\documentclass[sigconf]{acmart}

\newcommand{\adder}[0]{AD{$\Delta$}ER}

\usepackage[capitalize]{cleveref}
\usepackage{caption}
\usepackage{subcaption}
\usepackage{float}
\usepackage[htt]{hyphenat} 
\usepackage{array}
\usepackage{arydshln}

\usepackage{svg}
\usepackage{algorithm}
\usepackage{algpseudocode}
\usepackage{pifont}
\usepackage{lipsum}  
\usepackage{diagbox}
\usepackage{multirow}

\crefname{section}{Sec.}{Secs.}
\Crefname{section}{Section}{Sections}
\Crefname{table}{Table}{Tables}
\crefname{table}{Tab.}{Tabs.}

\usepackage{hyperref}
\hypersetup{
    colorlinks=true,
    linkcolor=blue,
    filecolor=magenta,      
    urlcolor=cyan,
    pdftitle={Overleaf Example},
    pdfpagemode=FullScreen,
    }
\usepackage{graphicx}

\AtBeginDocument{%
  }

\acmConference[]{}{}{}

\begin{document}

\title{\textit{adder-viz}: Real-Time Visualization Software for Transcoding Event Video }

\author{Andrew C. Freeman}
\email{andrew_freeman@baylor.edu}
\orcid{0000-0002-7927-8245}
\affiliation{%
  \institution{Baylor University}
  \city{Waco}
  \state{Texas}
  \country{USA}
}

\author{Luke Reinkensmeyer}
\email{luke_reinkensmeyer1@baylor.edu}
\orcid{0009-0004-4839-5168}
\affiliation{%
  \institution{Baylor University}
  \city{Waco}
  \state{Texas}
  \country{USA}
}


\begin{abstract}

  Recent years have brought about a surge in neuromorphic ``event'' video research, primarily targeting computer vision applications. Event video eschews video frames in favor of asynchronous, per-pixel intensity samples. While much work has focused on a handful of representations for specific event cameras, these representations have shown limitations in flexibility, speed, and compressibility. We previously proposed the unified \adder{} representation to address these concerns. This paper introduces numerous improvements to the \textit{adder-viz} software for visualizing real-time event transcode processes and applications in-the-loop. The MIT-licensed software is available from a centralized repository at \href{https://github.com/ac-freeman/adder-codec-rs}{https://github.com/ac-freeman/adder-codec-rs}.
\end{abstract}




\begin{teaserfigure}
    \centering
    \includegraphics[width=\linewidth]{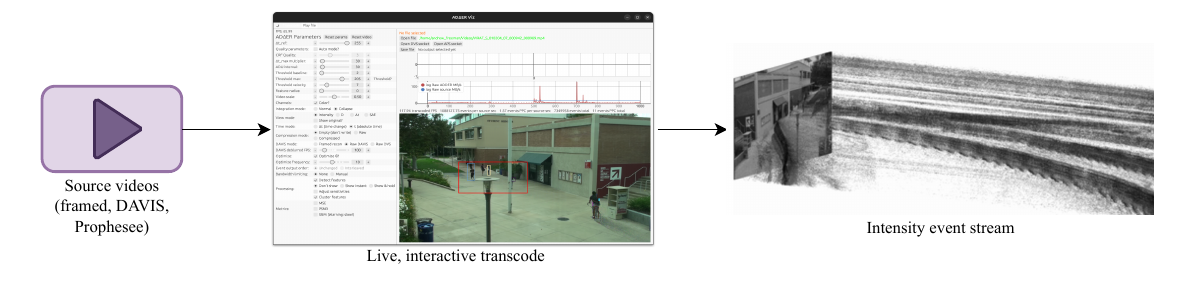}
    \caption{Usage of the \textit{adder-viz} interface. The small bounding boxes are obtained through FAST feature clustering. The red bounding box is a manually-drawn ROI. Regions outside this ROI (e.g., the cyclist) have lower sensitivity. These pixels produce few events, and experience temporal averaging of their intensities. }
    \label{fig:roi}
\end{teaserfigure}

\received{12 June 2025}
\received[accepted]{25 July 2025}
\received[revised]{20 August 2025}

\maketitle

\section{Introduction}\label{sec:intro}
Traditionally, video is structured as a sequence of discrete image frames. Recently, however, a novel video sensing paradigm has emerged which eschews video frames entirely. These ``event'' sensors aim to mimic the human vision system with asynchronous sensing, where each pixel has an independent, sparse data stream. While these cameras enable high-speed and high-dynamic-range sensing, researchers often revert to a framed representation of the event data for existing applications, or build bespoke applications for a particular camera’s event data type. At the same time, classical video systems have significant computational redundancy at the application layer, because pixel samples are repeated across frames in the uncompressed domain.

To address the shortcomings of existing systems, we previously introduced \textbf{A}ddress, \textbf{D}ecimation, $\Delta t$ \textbf{E}vent \textbf{R}epresentation (\textbf{\adder{}}, pronounced “adder''), a novel intermediate video representation and system framework \cite{freeman_accelerated_2024,freeman_asynchronous_2023}. The framework transcodes a variety of framed and event camera sources into a single event-based representation. This unified representation supports source-modeled lossy compression and backward compatibility with traditional frame-based applications.

Since the event video paradigm represents a dramatic shift from the mechanisms of classical video, we seek a simple, user-friendly program to increase the accessibility and reproducibility of this research avenue. Our \textit{adder-viz} software aims to fill this gap. While we previously introduced this software as a component of the broader \adder{} framework, it was severely limited in performance, reliability, accuracy, and features. In this work, we discuss the overall \adder{} framework, then introduce our recent advancements to our open-source software packages, including:

\begin{itemize}
    \item Improve the accuracy and clarity of transcode operations
    \item Remove dependency on unstable Rust features
    \item Expand interactive rate control applications with feature clustering and manual ROIs
    \item Add helper overlays for each transcoder setting
    \item Reduce the compiled binary size by $75\%$
    \item Increase live compression speed by up to $275\%$
    
\end{itemize}

\section{Background}

\subsection{Framed Video}

Since a traditional video is simply a sequence of discrete images, we see that an uncompressed video frame representation has high temporal and spatial redundancy. This redundancy can produce enormous data rates, especially at high frame rates and resolutions. Since the early days of video systems, researchers have sought to mitigate these rate concerns through compression, so that the bandwidth and storage capabilities of consumer devices can receive and process video with reasonable speed. The compression and application layers, however, are largely divorced in classical video systems, and compression performance does not correlate directly with application speed. This quality is most severe in systems where the video does achieve high temporal compression, such as surveillance and high-speed video. An application may incorporate a differencing mechanism to remove temporal redundancy, but such a scheme does not scale with changes to the input frame rate.


\subsection{Event Video}
Event video, on the other hand, detaches intensity samples from a strictly synchronous representation. Neuromorphic event cameras have recently received attention in computer vision research because they have microsecond temporal resolution, consume little power, and extremely high dynamic range (>120 dB) \cite{gallego_event-based_2022}. The fundamental advancement that makes these features possible is an asynchronous sensor circuit: whereas traditional sensors record the integrated value for all pixels at the same time, event-based sensors record an individual pixel's value only when it has met a certain threshold criterion.

The most common event camera leverages the Dynamic Vision System (DVS) \cite{lichtsteiner_128x128_2008, gallego_event-based_2022}. A DVS pixel records (or ``fires'') a tuple $\langle x, y, p, t\rangle$ at timestamp $t$ when it experiences a change in instantaneous log intensity that exceeds its global threshold, $\theta$ \cite{gallego_event-based_2022}. As a consequence of this change-based sensing, these cameras predominantly fire events that correspond to moving edges between areas of high contrast. Since static regions are thus not captured directly, researchers have sought to utilize multimodal sensing with DVS and frame-based cameras. The Dynamic and Active Vision System (DAVIS) combines a DVS and framed sensor on a single sensor package \cite{brandli_240_2014,brandli_real-time_2014}. The camera outputs separate data streams for frames and events, however, leaving sensor fusion as a downstream task. 

Various event camera manufacturers have their own proprietary data formats. Furthermore, many event-based applications are designed around a number of frame-based representations of the event data \cite{baldwin_time-ordered_2023,gehrig_end--end_2019,maqueda_event-based_2018,rebecq_real-time_2017,zhu_ev-flownet_2018}. This process inevitably quantizes the event timestamps, and it leaves any rate control mechanism up to a traditional frame-based encoder. Typical vision applications are designed around a single event format and representation, which makes real-world system integration difficult.

\subsection{\adder}
Our \adder{} system addresses some major weakness of both traditional framed video and existing event video representations. Chiefly, it transcodes video from a variety of sources into a \textbf{single event representation}. This transcode process supports variable loss of the raw representation, application in-the-loop support, and source-modeled compression.

In the transcoding process, an \adder{} event tuple is fired when a pixel's change in intensity accumulates to an amount past a certain threshold value. An event tuple $\langle x,y,c,D,t\rangle$ stores the spatial coordinates $(x)$ and $(y)$, the color channel $(c)$, the intensity threshold $(D)$, and the timestamp $(t)$. The threshold $D$ is determined by both the current pixel brightness and the time elapsed since the last event was fired \cite{freeman_asynchronous_2023}. $D$ can be  tuned further based on application-level demands, such as regional saliency \cite{freeman_accelerated_2024}. The intensity of a pixel is calculated by the equation $I = \frac{2^D}{\Delta t}$, where $\Delta t$ is the time elapsed since the pixel last fired an event. Unlike DVS events, which merely represent intensity \textit{change}, \adder{} events directly convey \textit{absolute intensity}.

\begin{figure}
    \centering
    \includegraphics[width=0.96\linewidth]{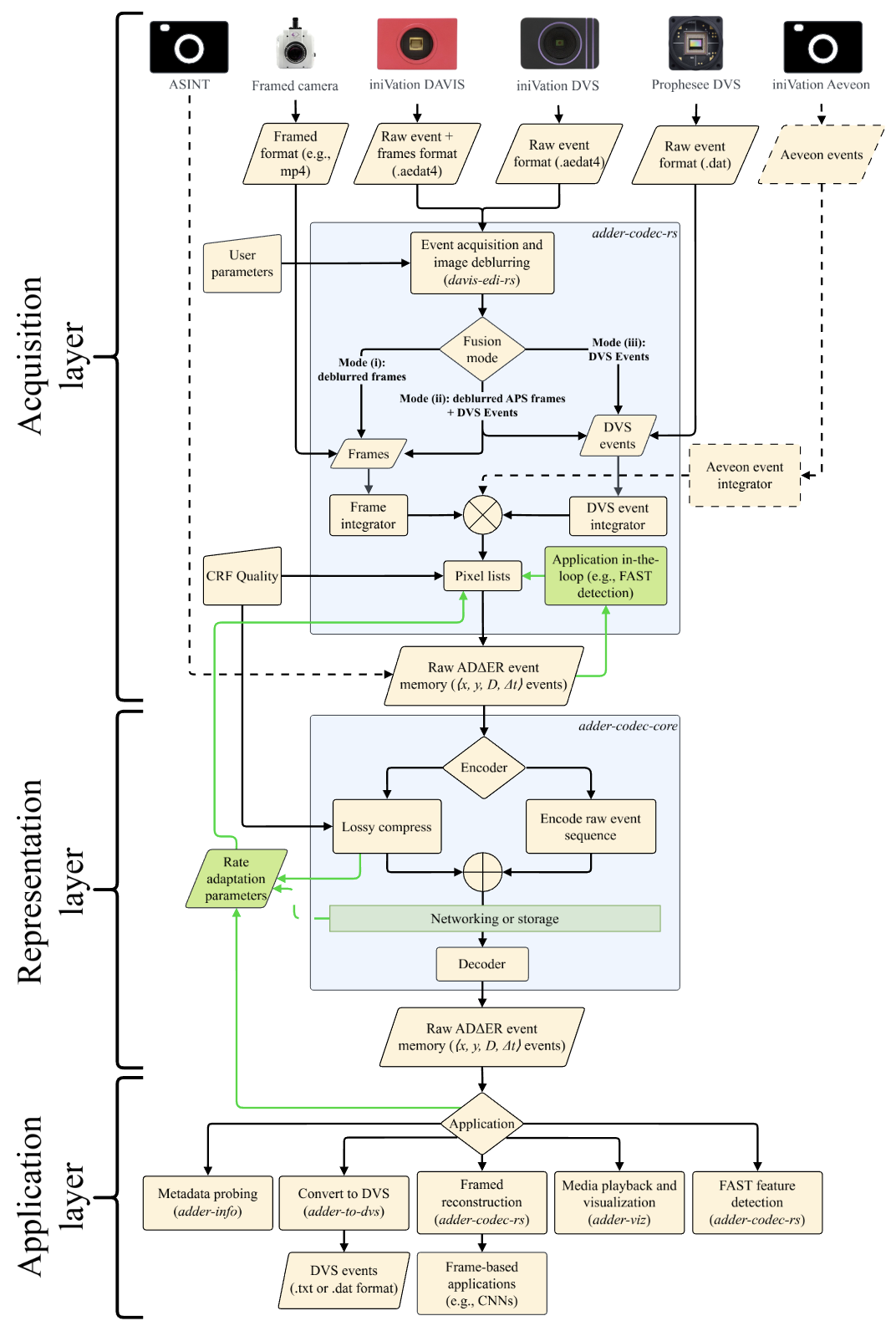}
    \caption[Detailed diagram of the \adder{} framework]{Detailed diagram of the three-layer \adder{} framework. Italicized names reflect the names of software packages in the Rust Package Registry. Dashed lines indicate future work for as yet unreleased cameras. Since multimodal sources can be transcoded to a single raw representation, we can have a simple source-modeled compression scheme. The representation supports bespoke event-based applications, while being backwards compatible with classical applications.}
    \label{fig:software_diagram}
\end{figure}

\textbf{Key advantages to \adder{} include the following:} it is source agnostic, so downstream applications may be transparently applied to both framed- and event-based video sources, including upcoming event cameras; it is trivial to produce a framed representation from \adder{}, so it maintains compatibility with a plethora of framed applications; it can dramatically reduce the raw (uncompressed) data rate of videos with high temporal redundancy, making vision applications faster \cite{freeman_accelerated_2024}; and, finally, it provides clear methods for introducing dynamic temporal loss for rate adaptation.

An example scenario where \adder{} can excel is computer vision analysis for surveillance systems. Other potential application areas include streaming systems for robotics, autonomous vehicles, and mixed reality. On a framed surveillance dataset, we found that \adder{} transcoding can reduce the amount of raw data by more than 90\% \cite{freeman_rethinking_2024}. If vision applications are adapted to operate on this temporally sparse representation, they can be made significantly faster. For example, we saw up to a 43\% speed-up in FAST feature detection in \cite{freeman_accelerated_2024}. If \adder{} transcoding occurs on-camera at the edge, then we can see faster transmission, reduced storage costs, and lower latency analysis. Furthermore, non-framed sensors such as a DAVIS camera may be integrated without requiring a separate processing pipeline. 

\section{Software Overview}

We first introduced the open-source \adder{} software suite in \cite{freeman_open_2024}. It consists of several libraries and applications written in Rust. We provide a modular interface for reading and transcoding data from multiple event cameras and framed video formats, to produce a raw \adder{} event stream. An \adder{} stream may be lossy compressed with our bespoke source-modeled scheme. We further provide applications which interface with our decoder utility to provide functionality such as command-line metadata inspection, framed reconstruction, playback and visualization, and feature detection. \cref{fig:software_diagram} illustrates how these software components are connected.

 Although the \adder{} software suite provides several libraries for creating and processing our event streams, our GUI program \textit{adder-viz} provides the most accessible starting point for researchers. Originally developed in late 2022, it provides researchers with a simple interface to experiment with various settings for \adder{} transcoding. A user can simply drag and drop a video file into the window to begin a transcode operation. The program visualizes the transcode by continuously updating an image display with the most recently fired event for each pixel. Transcoder settings are exposed with sliders and buttons, and the user can immediately observe the effect of changing each setting. \textit{adder-viz} further provides an interface for replaying saved \texttt{.adder} files.

Our software is available on \href{https://github.com/ac-freeman/adder-codec-rs}{GitHub} under the MIT license and distributed on the Rust Package Registry, \href{https://crates.io/crates/adder-codec-rs}{crates.io}. In our GitHub README, we provide user-friendly installation instructions, background reading, and a VirtualBox image. All source code documentation is available online on \href{docs.rs}{docs.rs}. According to metrics on \href{https://crates.io/crates/adder-codec-rs}{crates.io}, the various \adder{} packages have been collectively downloaded more than 150,000 times.

\section{Software Enhancements}
In the following, we describe in detail the major improvements made to our \adder{} software since our last major version in January 2024, as summarized in \cref{sec:intro}. Many of these improvements are brought directly to the \textit{adder-viz} program, while some are made to the underlying codec library.

\subsection{Usability}

Previously, we used the Bevy game engine \cite{noauthor_bevyenginebevy_2025} as our graphics back-end alongside the \textit{egui} front-end library. We found that the entity component system (ECS) of Bevy was overly complex for our purposes, and the numerous dependencies caused extremely slow compile times. We thus replaced Bevy with the lightweight \textit{eframe} renderer \cite{ernerfeldt_emilkegui_2025}, which allows developers to compile their changes in a matter of seconds. A positive side effect of this library change is that our new compiled binary is $75\%$ smaller than before.

Based on user feedback, we significantly streamlined the setup and usage of our program. Our prior version required the user to use the ``nightly'' branch of Rust if they wanted to perform source-modeled compression. This was a source of frustration, because the inherently unstable Rust branch would often break the program compilation. To solve this, we upstreamed improvements to the \textit{arithmetic-coding} library \cite{danieleades_danieleadesarithmetic-coding_2025} (which our codec depends on) that allow it to compile on the stable Rust branch. To make the program more approachable to new users, we also added a hover overlay to each transcoder setting. These overlays briefly describe what the settings do, and they give users the necessary context to search for further information in our documentation or associated papers. Finally, we added clearer instructions in our README for users to set up and install \textit{adder-viz} for the first time. These efforts help ensure that our software is accessible for the open-source community.

\subsection{Accuracy}

Through our own experiments with visualizing \adder{} transcode operations in \textit{adder-viz} we found a number of latent bugs that were dramatically impacting the accuracy of our produced representations.

\begin{figure}
\centering
\begin{subfigure}{.5\linewidth}
  \centering
  \includegraphics[width=\linewidth]{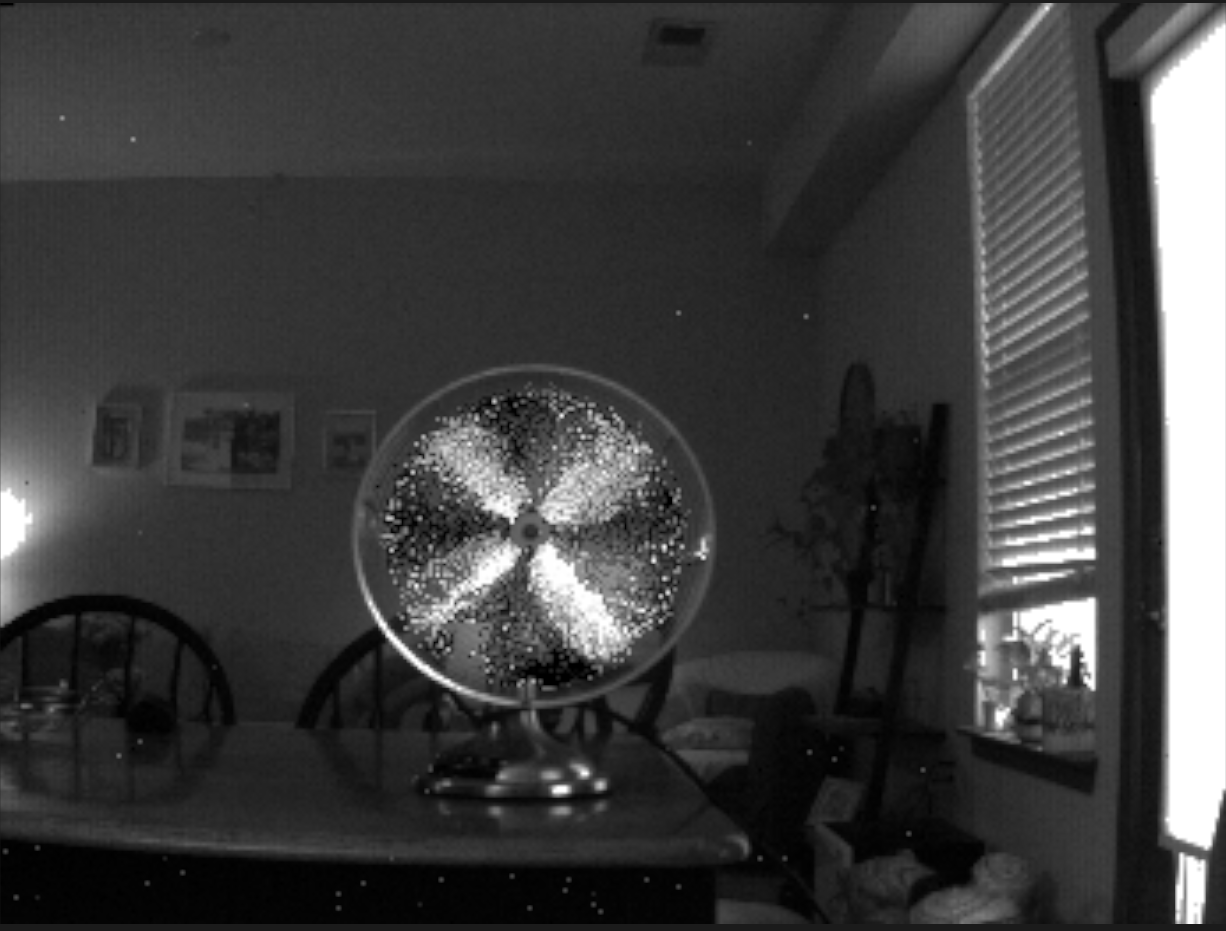}
  \caption{Old \cite{freeman_open_2024}}
  \label{fig:sub1}
\end{subfigure}%
\begin{subfigure}{.5\linewidth}
  \centering
  \includegraphics[width=\linewidth]{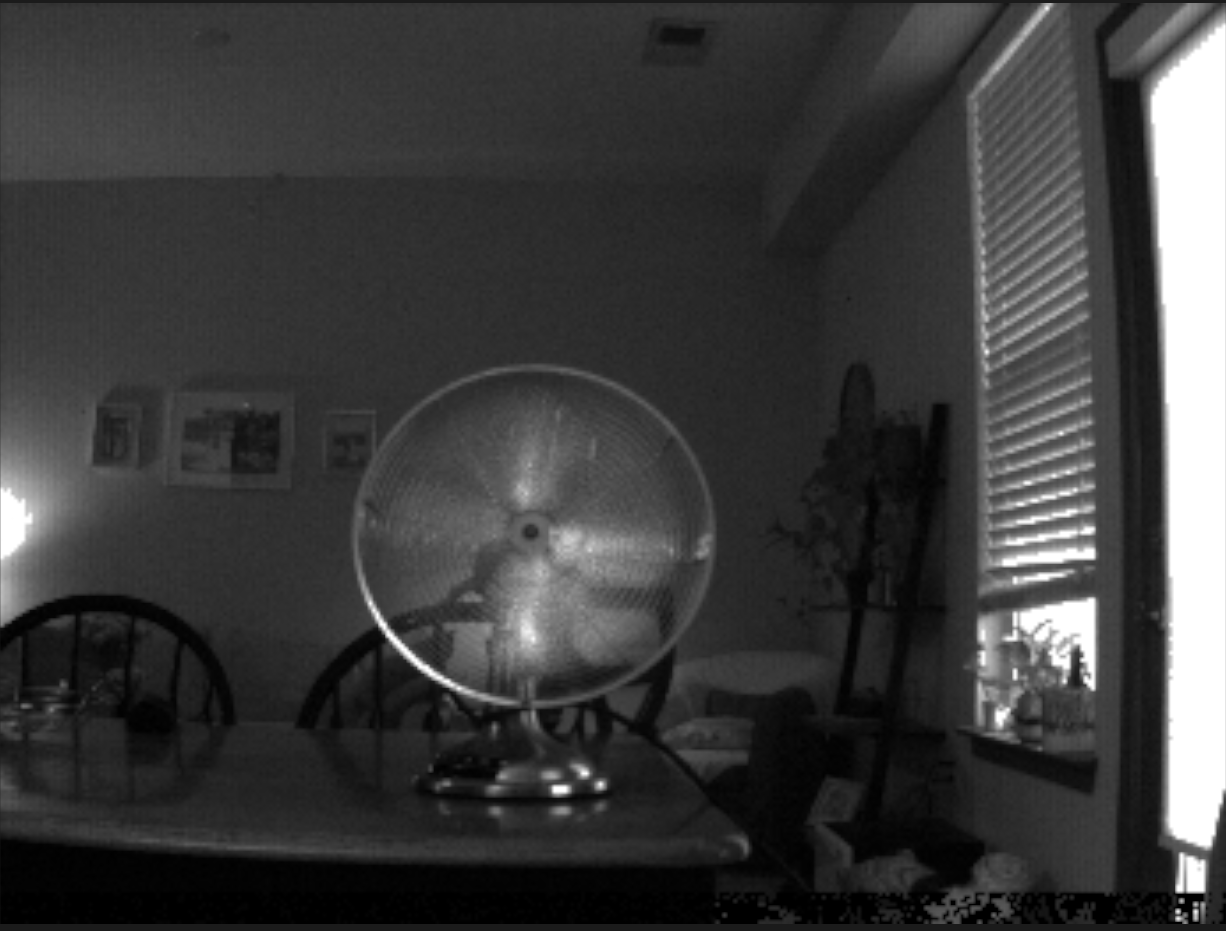}
  \caption{New (this work)}
  \label{fig:sub2}
\end{subfigure}
\caption{Qualitative example of the accuracy improvement to our DAVIS transcoder. Our events now have a more natural appearance. These images were recorded from reconstructed frames displayed in our \textit{adder-viz} event video player.}
\label{fig:fans}
\end{figure}

Firstly, we found that our transcoder did not output the correct data necessary to reconstruct a pixel value with zero intensity. Since our minimum $D$ is 0, we do not fire an event until accumulating $2^0 = 1$ intensity unit. If a pixel has zero intensity for a sustained period, then, it will not produce the necessary events. To solve this, we defined a special case with a $D$ constant for such intensities. Secondly, we corrected an algorithmic mistake in our DAVIS transcoder. With our correction, the output \adder{} events have higher accuracy and better temporal coherence. We illustrate this difference in \cref{fig:fans}.

\subsection{Interactive Applications}

We extended our FAST feature detection \cite{rosten_fusing_2005} application by implementing the Density-Based Spatial Clustering of Applications with Noise (DBSCAN) \cite{ester_density-based_1996} algorithm to cluster the FAST features. We perform this clustering during \adder{} transcode, operating at regular intervals (e.g., 30 times per second for event-based source, or matching the input rate of framed sources) . These feature clusters then serve as prototypical regions for moving object detection, without the use of a learning-based method. We tuned the feature detection settings to avoid noisy regions and small clusters, instead prioritizing large clusters with several features. We provide a toggle to visualize these cluster bounding boxes on the live display in \textit{adder-viz}, as shown in \cref{fig:roi}.

We also added a tool for a user to manually add a region of interest (ROI) during a transcode operation. The user simply clicks and drags within the \textit{adder-viz} display to create an ROI. Pixels within the ROI will be set to have a high sensitivity, while pixels outside the ROI proceed as normal. By creating an ROI in combination with a high CRF value, the user can visualize the unique rate properties of \adder: pixels with high sensitivity will produce more frequent events and a higher visual quality, and pixels with low sensitivity will produce few events. We illustrate this saliency-based rate control in \cref{fig:roi}.

\subsection{Speed}

Previously, our UI updates were performed on the same thread as the main transcoder process. Although convenient, the shortcoming of this design was severe lag when the transcoder could not process the incoming data in real time (e.g., when transcoding high-resolution video). Now, the UI is handled in a dedicated thread, and the transcoder uses a concurrent data structure to push updated visualization frames to the UI.

\begin{table}
    \centering
    \begin{tabular}{cc|ll|ll}
        & & \multicolumn{2}{c|}{Raw events} & \multicolumn{2}{c}{Lossy compression} \\
          
         \multicolumn{2}{c|}{Res.} & Grayscale & Color & Grayscale & Color \\
        \hline \parbox[t]{1mm}{\multirow{3}{*}{\rotatebox[origin=c]{90}{\cite{freeman_open_2024}}}} 
        & SD   & 60  & 60  &  60  & 52 \\
        & HD  &  53 &  35 &  39  & 14 \\
        & FHD &  35 &  16 &  17  & 4 \\
    \hline \parbox[t]{1mm}{\multirow{3}{*}{\rotatebox[origin=c]{90}{\textbf{New}}}} 
        & SD   & 150 (+150\%) & 90 (+50\%)  & 129  (+115\%) & 63 (+21\%)\\
        & HD  & 85 (+60\%) & 44 (+26\%) & 62 (+59\%)  & 32 (+128\%)\\
        & FHD & 49 (+40\%) & 24 (+50\%) &  38 (+124\%) & 15 (+275\%)\\
    \end{tabular}
    \caption{Average frames per second at which the ToS framed video sequence can be transcoded within \textit{adder-viz}. 
    A CRF value of 3 (the default) was used for these experiments.}
    \label{tab:perf_framed}
    \vspace{-\baselineskip}
\end{table}

Additionally, our source-modeled compression mechanism frequently proved to be a bottleneck: when a group of events was compressed, the transcode and UI processes would freeze. We therefore moved the compressor to its own concurrent process, where it receives events through message passing from the transcoder and compresses them asynchronously. Due to the computational complexity of the compression, we still experience slower performance when compression is enabled; however, the slowdown is now consistent across time, and does not cause noticeable pauses during transcoding.

We compared the new version of \textit{adder-viz} to that of our prior work \cite{freeman_open_2024}. For input, we used the Tears of Steel (Tos) \cite{roosendaal_about_2012} sequence at various resolutions, both in grayscale and color, and both with and without our bespoke compression. The ToS sequence is a live-action movie with computer-generated video effects. We further validated the speed of transcoding and compressing full-motion DAVIS event camera video captured from a moving car. Our system has an Intel Core i9-14900K with 64 GB of RAM, running Ubuntu 24.04. We summarize our results in \cref{tab:perf_framed}.

In total, our optimizations yield a tremendous improvement in the transcoder visualization speed. We see that the maximum framed transcoder speed we encountered before was merely 60 FPS, matching the refresh rate of our display. With the transcoder process divorced from the rendering thread, we now see much higher frame rates at low resolutions. Even in cases where we were not limited by the refresh rate or compressor lag, we still see improvements of 26\% or more. In the best case, we observe an improvement of \textbf{275\%} when transcoding and compressing color Full-HD video. When transcoding our DAVIS video with frame-event fusion (mode (ii) of \cref{fig:software_diagram}), we observe a maximum latency of 29 ms when compression is enabled. For transcoding only the DVS events (mode (iii) of \cref{fig:software_diagram}), we observe a maximum latency of 25 ms with compression enabled. Finally, we found that the feature detection, clustering, and bounding box display systems had no measurable impact on performance with up to 25 bounding boxes displayed at once.

\section{Conclusion}

\adder{} presents a unique alternative to traditional frame- and event-based sensing: it nonuniformly encodes intensity samples based on their rate of change, and the raw bitrate adjusts dynamically to the motion content. By increasing the temporal sparsity of the video representation, we can highly compress temporally stable regions without degrading the image quality. Our open-source software lowers barriers for new researchers to investigate event video. With further development of the compression and application techniques, we may see dramatic efficiency gains for large-scale video analysis systems.

\begin{acks}
Some text and images throughout were adapted from the first author's recent dissertation \cite{freeman_rethinking_2024}.
\end{acks}

\bibliographystyle{ACM-Reference-Format}
\bibliography{references}


\begin{thebibliography}{19}


\ifx \showCODEN    \undefined \def \showCODEN     #1{\unskip}     \fi
\ifx \showDOI      \undefined \def \showDOI       #1{#1}\fi
\ifx \showISBNx    \undefined \def \showISBNx     #1{\unskip}     \fi
\ifx \showISBNxiii \undefined \def \showISBNxiii  #1{\unskip}     \fi
\ifx \showISSN     \undefined \def \showISSN      #1{\unskip}     \fi
\ifx \showLCCN     \undefined \def \showLCCN      #1{\unskip}     \fi
\ifx \shownote     \undefined \def \shownote      #1{#1}          \fi
\ifx \showarticletitle \undefined \def \showarticletitle #1{#1}   \fi
\ifx \showURL      \undefined \def \showURL       {\relax}        \fi
\providecommand\bibfield[2]{#2}
\providecommand\bibinfo[2]{#2}
\providecommand\natexlab[1]{#1}
\providecommand\showeprint[2][]{arXiv:#2}

\bibitem[noa(2025)]%
        {noauthor_bevyenginebevy_2025}
 \bibinfo{year}{2025}\natexlab{}.
\newblock \bibinfo{title}{bevyengine/bevy}.
\newblock
\newblock
\urldef\tempurl%
\url{https://github.com/bevyengine/bevy}
\showURL{%
\tempurl}
\newblock
\shownote{original-date: 2020-01-18T21:13:55Z}.


\bibitem[Baldwin et~al\mbox{.}(2023)]%
        {baldwin_time-ordered_2023}
\bibfield{author}{\bibinfo{person}{R.~Wes Baldwin}, \bibinfo{person}{Ruixu Liu}, \bibinfo{person}{Mohammed Almatrafi}, \bibinfo{person}{Vijayan Asari}, {and} \bibinfo{person}{Keigo Hirakawa}.} \bibinfo{year}{2023}\natexlab{}.
\newblock \showarticletitle{Time-{Ordered} {Recent} {Event} ({TORE}) {Volumes} for {Event} {Cameras}}.
\newblock \bibinfo{journal}{\emph{IEEE Transactions on Pattern Analysis and Machine Intelligence}} \bibinfo{volume}{45}, \bibinfo{number}{2} (\bibinfo{date}{Feb.} \bibinfo{year}{2023}), \bibinfo{pages}{2519--2532}.
\newblock
\showISSN{1939-3539}
\urldef\tempurl%
\url{https://doi.org/10.1109/TPAMI.2022.3172212}
\showDOI{\tempurl}


\bibitem[Brandli et~al\mbox{.}(2014a)]%
        {brandli_240_2014}
\bibfield{author}{\bibinfo{person}{Christian Brandli}, \bibinfo{person}{Raphael Berner}, \bibinfo{person}{Minhao Yang}, \bibinfo{person}{Shih-Chii Liu}, {and} \bibinfo{person}{Tobi Delbruck}.} \bibinfo{year}{2014}\natexlab{a}.
\newblock \showarticletitle{A 240 × 180 130 {dB} 3 µs {Latency} {Global} {Shutter} {Spatiotemporal} {Vision} {Sensor}}.
\newblock \bibinfo{journal}{\emph{IEEE Journal of Solid-State Circuits}} \bibinfo{volume}{49}, \bibinfo{number}{10} (\bibinfo{date}{Oct.} \bibinfo{year}{2014}), \bibinfo{pages}{2333--2341}.
\newblock
\showISSN{1558-173X}
\urldef\tempurl%
\url{https://doi.org/10.1109/JSSC.2014.2342715}
\showDOI{\tempurl}


\bibitem[Brandli et~al\mbox{.}(2014b)]%
        {brandli_real-time_2014}
\bibfield{author}{\bibinfo{person}{Christian Brandli}, \bibinfo{person}{Lorenz Muller}, {and} \bibinfo{person}{Tobi Delbruck}.} \bibinfo{year}{2014}\natexlab{b}.
\newblock \showarticletitle{Real-time, high-speed video decompression using a frame- and event-based {DAVIS} sensor}. In \bibinfo{booktitle}{\emph{2014 {IEEE} {International} {Symposium} on {Circuits} and {Systems} ({ISCAS})}}. \bibinfo{pages}{686--689}.
\newblock
\urldef\tempurl%
\url{https://doi.org/10.1109/ISCAS.2014.6865228}
\showDOI{\tempurl}
\newblock
\shownote{ISSN: 2158-1525}.


\bibitem[danieleades(2025)]%
        {danieleades_danieleadesarithmetic-coding_2025}
\bibfield{author}{\bibinfo{person}{danieleades}.} \bibinfo{year}{2025}\natexlab{}.
\newblock \bibinfo{title}{danieleades/arithmetic-coding}.
\newblock
\newblock
\urldef\tempurl%
\url{https://github.com/danieleades/arithmetic-coding}
\showURL{%
\tempurl}
\newblock
\shownote{original-date: 2022-03-16T17:20:54Z}.


\bibitem[Ernerfeldt(2025)]%
        {ernerfeldt_emilkegui_2025}
\bibfield{author}{\bibinfo{person}{Emil Ernerfeldt}.} \bibinfo{year}{2025}\natexlab{}.
\newblock \bibinfo{title}{emilk/egui}.
\newblock
\newblock
\urldef\tempurl%
\url{https://github.com/emilk/egui}
\showURL{%
\tempurl}
\newblock
\shownote{original-date: 2019-01-13T15:39:15Z}.


\bibitem[Ester et~al\mbox{.}(1996)]%
        {ester_density-based_1996}
\bibfield{author}{\bibinfo{person}{Martin Ester}, \bibinfo{person}{Hans-Peter Kriegel}, \bibinfo{person}{Jörg Sander}, {and} \bibinfo{person}{Xiaowei Xu}.} \bibinfo{year}{1996}\natexlab{}.
\newblock \showarticletitle{A density-based algorithm for discovering clusters in large spatial databases with noise}. In \bibinfo{booktitle}{\emph{Proceedings of the {Second} {International} {Conference} on {Knowledge} {Discovery} and {Data} {Mining}}} \emph{(\bibinfo{series}{{KDD}'96})}. \bibinfo{publisher}{AAAI Press}, \bibinfo{address}{Portland, Oregon}, \bibinfo{pages}{226--231}.
\newblock


\bibitem[Freeman(2024a)]%
        {freeman_rethinking_2024}
\bibfield{author}{\bibinfo{person}{Andrew Freeman}.} \bibinfo{year}{2024}\natexlab{a}.
\newblock \emph{\bibinfo{title}{Rethinking {Video} with a {Universal} {Event}-{Based} {Representation}}}.
\newblock \bibinfo{thesistype}{Ph.\,D. Dissertation}. \bibinfo{school}{The University of North Carolina at Chapel Hill University Libraries}.
\newblock
\urldef\tempurl%
\url{https://doi.org/10.17615/5BSV-BZ25}
\showDOI{\tempurl}


\bibitem[Freeman(2024b)]%
        {freeman_open_2024}
\bibfield{author}{\bibinfo{person}{Andrew~C. Freeman}.} \bibinfo{year}{2024}\natexlab{b}.
\newblock \showarticletitle{An {Open} {Software} {Suite} for {Event}-{Based} {Video}}. In \bibinfo{booktitle}{\emph{Proceedings of the {ACM} {Multimedia} {Systems} {Conference} 2024}}. \bibinfo{publisher}{ACM}, \bibinfo{address}{Bari Italy}, \bibinfo{pages}{271--277}.
\newblock
\showISBNx{979-8-4007-0412-3}
\urldef\tempurl%
\url{https://doi.org/10.1145/3625468.3652169}
\showDOI{\tempurl}


\bibitem[Freeman et~al\mbox{.}(2024)]%
        {freeman_accelerated_2024}
\bibfield{author}{\bibinfo{person}{Andrew~C. Freeman}, \bibinfo{person}{Ketan Mayer-Patel}, {and} \bibinfo{person}{Montek Singh}.} \bibinfo{year}{2024}\natexlab{}.
\newblock \showarticletitle{Accelerated {Event}-{Based} {Feature} {Detection} and {Compression} for {Surveillance} {Video} {Systems}}. In \bibinfo{booktitle}{\emph{Proceedings of the 15th {ACM} {Multimedia} {Systems} {Conference}}} \emph{(\bibinfo{series}{{MMSys} '24})}. \bibinfo{publisher}{Association for Computing Machinery}, \bibinfo{address}{New York, NY, USA}, \bibinfo{pages}{132--143}.
\newblock
\showISBNx{979-8-4007-0412-3}
\urldef\tempurl%
\url{https://doi.org/10.1145/3625468.3647618}
\showDOI{\tempurl}


\bibitem[Freeman et~al\mbox{.}(2023)]%
        {freeman_asynchronous_2023}
\bibfield{author}{\bibinfo{person}{Andrew~C. Freeman}, \bibinfo{person}{Montek Singh}, {and} \bibinfo{person}{Ketan Mayer-Patel}.} \bibinfo{year}{2023}\natexlab{}.
\newblock \showarticletitle{An {Asynchronous} {Intensity} {Representation} for {Framed} and {Event} {Video} {Sources}}. In \bibinfo{booktitle}{\emph{Proceedings of the 14th {ACM} {Multimedia} {Systems} {Conference}}}. \bibinfo{publisher}{ACM}, \bibinfo{address}{Vancouver BC Canada}, \bibinfo{pages}{74--85}.
\newblock
\showISBNx{979-8-4007-0148-1}
\urldef\tempurl%
\url{https://doi.org/10.1145/3587819.3590969}
\showDOI{\tempurl}


\bibitem[Gallego et~al\mbox{.}(2022)]%
        {gallego_event-based_2022}
\bibfield{author}{\bibinfo{person}{Guillermo Gallego}, \bibinfo{person}{Tobi Delbrück}, \bibinfo{person}{Garrick Orchard}, \bibinfo{person}{Chiara Bartolozzi}, \bibinfo{person}{Brian Taba}, \bibinfo{person}{Andrea Censi}, \bibinfo{person}{Stefan Leutenegger}, \bibinfo{person}{Andrew~J. Davison}, \bibinfo{person}{Jörg Conradt}, \bibinfo{person}{Kostas Daniilidis}, {and} \bibinfo{person}{Davide Scaramuzza}.} \bibinfo{year}{2022}\natexlab{}.
\newblock \showarticletitle{Event-{Based} {Vision}: {A} {Survey}}.
\newblock \bibinfo{journal}{\emph{IEEE Transactions on Pattern Analysis and Machine Intelligence}} \bibinfo{volume}{44}, \bibinfo{number}{01} (\bibinfo{date}{Jan.} \bibinfo{year}{2022}), \bibinfo{pages}{154--180}.
\newblock
\showISSN{0162-8828}
\urldef\tempurl%
\url{https://doi.org/10.1109/TPAMI.2020.3008413}
\showDOI{\tempurl}
\newblock
\shownote{Publisher: IEEE Computer Society}.


\bibitem[Gehrig et~al\mbox{.}(2019)]%
        {gehrig_end--end_2019}
\bibfield{author}{\bibinfo{person}{Daniel Gehrig}, \bibinfo{person}{Antonio Loquercio}, \bibinfo{person}{Konstantinos Derpanis}, {and} \bibinfo{person}{Davide Scaramuzza}.} \bibinfo{year}{2019}\natexlab{}.
\newblock \showarticletitle{End-to-{End} {Learning} of {Representations} for {Asynchronous} {Event}-{Based} {Data}}. In \bibinfo{booktitle}{\emph{2019 {IEEE}/{CVF} {International} {Conference} on {Computer} {Vision} ({ICCV})}}. \bibinfo{publisher}{IEEE}, \bibinfo{address}{Seoul, Korea (South)}, \bibinfo{pages}{5632--5642}.
\newblock
\showISBNx{978-1-7281-4803-8}
\urldef\tempurl%
\url{https://doi.org/10.1109/ICCV.2019.00573}
\showDOI{\tempurl}


\bibitem[Lichtsteiner et~al\mbox{.}(2008)]%
        {lichtsteiner_128x128_2008}
\bibfield{author}{\bibinfo{person}{Patrick Lichtsteiner}, \bibinfo{person}{Christoph Posch}, {and} \bibinfo{person}{Tobi Delbruck}.} \bibinfo{year}{2008}\natexlab{}.
\newblock \showarticletitle{A 128x128 120 {dB} 15 microsend {Latency} {Asynchronous} {Temporal} {Contrast} {Vision} {Sensor}}.
\newblock \bibinfo{journal}{\emph{IEEE Journal of Solid-State Circuits}} \bibinfo{volume}{43}, \bibinfo{number}{2} (\bibinfo{date}{Feb.} \bibinfo{year}{2008}), \bibinfo{pages}{566--576}.
\newblock
\showISSN{1558-173X}
\urldef\tempurl%
\url{https://doi.org/10.1109/JSSC.2007.914337}
\showDOI{\tempurl}
\newblock
\shownote{Conference Name: IEEE Journal of Solid-State Circuits}.


\bibitem[Maqueda et~al\mbox{.}(2018)]%
        {maqueda_event-based_2018}
\bibfield{author}{\bibinfo{person}{Ana~I. Maqueda}, \bibinfo{person}{Antonio Loquercio}, \bibinfo{person}{Guillermo Gallego}, \bibinfo{person}{Narciso Garcia}, {and} \bibinfo{person}{Davide Scaramuzza}.} \bibinfo{year}{2018}\natexlab{}.
\newblock \showarticletitle{Event-{Based} {Vision} {Meets} {Deep} {Learning} on {Steering} {Prediction} for {Self}-{Driving} {Cars}}. In \bibinfo{booktitle}{\emph{2018 {IEEE}/{CVF} {Conference} on {Computer} {Vision} and {Pattern} {Recognition}}}. \bibinfo{publisher}{IEEE}, \bibinfo{address}{Salt Lake City, UT}, \bibinfo{pages}{5419--5427}.
\newblock
\showISBNx{978-1-5386-6420-9}
\urldef\tempurl%
\url{https://doi.org/10.1109/CVPR.2018.00568}
\showDOI{\tempurl}


\bibitem[Rebecq et~al\mbox{.}(2017)]%
        {rebecq_real-time_2017}
\bibfield{author}{\bibinfo{person}{Henri Rebecq}, \bibinfo{person}{Timo Horstschaefer}, {and} \bibinfo{person}{Davide Scaramuzza}.} \bibinfo{year}{2017}\natexlab{}.
\newblock \showarticletitle{Real-time {Visual}-{Inertial} {Odometry} for {Event} {Cameras} using {Keyframe}-based {Nonlinear} {Optimization}}. In \bibinfo{booktitle}{\emph{Proceedings of the {British} {Machine} {Vision} {Conference} 2017}}. \bibinfo{publisher}{British Machine Vision Association}, \bibinfo{address}{London, UK}, \bibinfo{pages}{16}.
\newblock
\showISBNx{978-1-901725-60-5}
\urldef\tempurl%
\url{https://doi.org/10.5244/C.31.16}
\showDOI{\tempurl}


\bibitem[Roosendaal(2012)]%
        {roosendaal_about_2012}
\bibfield{author}{\bibinfo{person}{Ton Roosendaal}.} \bibinfo{year}{2012}\natexlab{}.
\newblock \bibinfo{title}{About {\textbar} {Tears} of {Steel}}.
\newblock
\newblock
\urldef\tempurl%
\url{https://mango.blender.org/about/}
\showURL{%
\tempurl}


\bibitem[Rosten and Drummond(2005)]%
        {rosten_fusing_2005}
\bibfield{author}{\bibinfo{person}{E. Rosten} {and} \bibinfo{person}{T. Drummond}.} \bibinfo{year}{2005}\natexlab{}.
\newblock \showarticletitle{Fusing points and lines for high performance tracking}. In \bibinfo{booktitle}{\emph{Tenth {IEEE} {International} {Conference} on {Computer} {Vision} ({ICCV}'05) {Volume} 1}}, Vol.~\bibinfo{volume}{2}. \bibinfo{pages}{1508--1515 Vol. 2}.
\newblock
\urldef\tempurl%
\url{https://doi.org/10.1109/ICCV.2005.104}
\showDOI{\tempurl}
\newblock
\shownote{ISSN: 2380-7504}.


\bibitem[Zhu et~al\mbox{.}(2018)]%
        {zhu_ev-flownet_2018}
\bibfield{author}{\bibinfo{person}{Alex~Zihao Zhu}, \bibinfo{person}{Liangzhe Yuan}, \bibinfo{person}{Kenneth Chaney}, {and} \bibinfo{person}{Kostas Daniilidis}.} \bibinfo{year}{2018}\natexlab{}.
\newblock \showarticletitle{{EV}-{FlowNet}: {Self}-{Supervised} {Optical} {Flow} {Estimation} for {Event}-based {Cameras}}. In \bibinfo{booktitle}{\emph{Robotics: {Science} and {Systems} {XIV}}}.
\newblock
\urldef\tempurl%
\url{https://doi.org/10.15607/RSS.2018.XIV.062}
\showDOI{\tempurl}
\newblock
\shownote{arXiv:1802.06898 [cs]}.


\end{thebibliography}










\end{document}